\documentclass[fleqn,usenatbib]{mnras}
\usepackage{newtxtext,newtxmath}
\usepackage[T1]{fontenc}
\usepackage{ae,aecompl}

\usepackage{graphicx}	
\usepackage{amsmath}	
\usepackage{amssymb}	
\usepackage{xcolor}     
\usepackage{siunitx}

\title[Lopsidedness from dynamical friction]{Tidal forces from the wake of dynamical friction: warps, lopsidedness and kinematic misalignment}

\author[Kipper et al.]{
Rain Kipper$^{1,2}$\thanks{E-mail: rain.kipper@ut.ee}, Mar\'ia Benito$^{3}$, Peeter Tenjes$^{1}$, Elmo Tempel$^{1}$, Roberto de Propris$^{2}$
\\
$^{1}$Tartu Observatory, University of Tartu, Observatooriumi 1, 61602 T\~oravere, Estonia\\
$^2$FINCA, University of Turku, Vesilinnantie 5, 20014, Turku, Finland\\
$^{3}$National Institute of Chemical Physics and Biophysics, R\"avala 10, Tallinn 10143, Estonia\\
}

\date{Accepted XXX. Received YYY; in original form ZZZ}

\pubyear{2015}

\begin{document}
\label{firstpage}
\pagerange{\pageref{firstpage}--\pageref{lastpage}}
\maketitle

\begin{abstract}
 A galaxy moving through a background of dark matter particles, induces an overdensity of these particles or a wake behind it. The back reaction of this wake on the galaxy is a force field that can be decomposed into an effective deceleration (called dynamical friction) and a tidal field. In this paper we determine the tidal forces, thus generated on the galaxy, and the resulting observables, which are shown to be warps, lopsidedness and/or kinematic-photometric position angle misalignments. We estimate the magnitude of the tidal-like effects needed to reproduce the observed warp and lopsidedness on the isolated galaxy IC 2487. Within a realistic range of dark matter distribution properties the observed warped and lopsided kinematical properties of IC 2487 is possible to reproduce (the background medium of dark matter particles has a velocity dispersion of $\lesssim 80\,{\rm km\,s^{-1}}$ and the density $10^4-10^5~{\rm M_\odot\,kpc^{-3}}$, more likely at the lower end). We conclude that the proposed mechanism can generate warps, lopsidedness and misalignments observed in isolated galaxies or galaxies in loose groups. The method can be used also to constrain dark matter spatial and velocity distribution properties.
\end{abstract}


\section{Introduction}
Asymmetries in the mass and light distribution of galaxies are their well-known characteristics. 
For instance, \citet{SanchezSaavedra:2003} found that 54 per cent of their sample of 276 edge-on galaxies are warped.
In studies by \citet{richter:1994} and \citet{haynes:1998} about 50 per cent of galaxies have lopsided gas distributions\footnote{{ The gas and/or stellar component of a lopsided galaxy extend further out on one side of the galaxy with respect to the other side. This asymmetry is measured by the $m=1$ Fourier mode of the surface density.}};
\citet{Zaritsky:97} and \citet{Rix:1995hn} showed that 30 per cent of their sample of nearly face-on spiral galaxies have lopsided stellar discs. Thus, this kind of asymmetry is a wide spread and well-established property of galaxies. However, its origin and evolution is not theoretically fully understood.

As a galaxy moves through a background medium of dark matter (DM) particles, its gravitational potential induces a density wake in the surrounding medium. This DM wake generates a force field in the galaxy which can be decomposed into an effective deceleration of the galaxy, usually referred as dynamical friction (DF), and a tidal field \citep{Mulder1983}. This work studies whether this tidal field is able to cause asymmetries, such as lopsidedness {and/or warps}, in galaxies.
There is not an unique mechanism responsible for lopsided kinematics, lopsided mass distribution in galaxies { or warp formation}. Rather, several mechanisms can be invoked { \citep{Mapelli:2008, Jog_2009, 2006MNRAS.370....2S, 2009ApJ...703.2068D}}. Lopsided galaxies have been observed in clusters and groups of galaxies, as well as in isolated galaxies. By quantifying the correlation of lopsidedness with any of galactic property, we might better understand the mechanisms responsible for it.
In particular, lopsidedness does not correlate with the presence of a companion \citep{Bournaud:2005} and it has only slight correlation with the internal structure of the galaxy, e.g. intensity of spiral structure \citep{Zaritsky:2013}. 
Also, lopsidedness is higher in the outer parts of galaxies { and the phase of the first Fourier component of the surface density remains constant with radius \citep{2011A&A...530A..30V}, which suggests that lopsidedness is a global feature \citep{1997ApJ...488..642J}.}
In the following we briefly describe the main explanations for lopsidedness and their assessments. For a detailed review, we refer the interested reader to \citet{Jog_2009}. 

Encounters with nearby companions including tidal forces imparted by flyby galaxies \citep{Mapelli:2008} or minor mergers \citep{Zaritsky:97} can induce lopsidedness both on the inner kinematics and mass distribution of galaxies. This mechanism is disfavoured in the case of isolated galaxies due to the lack of companions. { However, for isolated galaxies it can not be {ruled out for a long-lived lopsided mode} \citep{2011A&A...530A..30V}.}

Gas accretion from cosmological filaments in a particular direction causes elongation of the mass distribution in this direction and, thus, it can cause lopsidedness. This mechanism leaves a signature in the kinematics since the tip of the asymmetry has motions equable to the escape velocity. However, lopsidedness in the kinematics is observed at the scale of rotational velocities. 
In addition, gas accretion does not produce noticeable radial velocity differences, as it would be expected, for instance, in the case of a recent infall \citep{Baldwin:1980}. Hence this explanation generally does not satisfy well, although, in some galaxies their asymmetries can be explained by a strong and constant accretion \citep{Bournaud:2005}. 

Another viable explanation for the lopsided geometry of a galaxy is that its DM halo is asymmetric. Thus, the disc's response to the halo potential causes the lopsided behaviour of the disc { \citep{1997ApJ...488..642J, Zaritsky:2013}}.
At present, the DM halo asymmetry explanation seems one of the most promising explanations to lopsidedness since it can match observations well \citep{Jog_2009} although the stability of the lopsidedness is a concern{, and the lifetime of the lopsided mode and its origin are correlated. If lopsidedness is treated as a purely kinematic feature, it winds up in a time scale which is less than a Gyr \citep{Baldwin:1980}. However, the disc's lopsidedness has been shown to be long-lived, with a lifetime larger than a Hubble time, if the pattern speed of the lopsided mode is small \citep{2002ApJ...568..190I, 2007MNRAS.382..419S}.} {\cite{2007MNRAS.382..419S} demonstrated this using a simulation of a purely exponential disk with no bulge or dark halo components.} 

{Several mechanisms have been invoked to explain warps. Proposed mechanisms include the misalignment between the disc's rotation axis and the principal axis of a non-spherical DM halo \citep{1988MNRAS.234..873S, 2009ApJ...703.2068D}, gas accretion \citep{1989MNRAS.237..785O, 2006MNRAS.370....2S}, or interactions with companions or satellites \citep{2006ApJ...641L..33W}. The stability of warps is also a concern -- on average, they tend to wear off in simulations in a timescale less than a Gyr \citep{semczuk2020tidally}.}

In this paper we propose that the DM wake responsible for DF further induces tidal-like effects that might cause asymmetry of the intrinsic dark halo of a galaxy, thus resulting in galaxy lopsidedness and/or other asymmetries such as warps. DF is a drag force that slows down a massive object by exchanging momentum with particles unbound to the object itself. This effective force is created by a collection of particles which have been gravitationally focused behind the object, thus, acting as a density wave which interacts with the massive object gravitationally. {Galaxies moving through the extended DM halos of clusters or groups of galaxies experience DF. For instance, DF pulls the brightest cluster galaxies (BCG) towards the centres of clusters (e.g. \cite{Zabludoff:1990}). Frictional effects are also experienced by objects orbiting withing galaxies. An example is the Large Magellanic Cloud (LMC) which is the most massive satellite of the Milky Way (MW). Recent studies have shown that the gravitational interaction of the MW and the LMC should have left observable imprints in the kinematics of halo stars in our Galaxy \citep{Garavito:2019, Petersen+20, Erkal+20}.}

As a galaxy moves through an external field of DM particles {which are not bound to the galaxy itself}, the particles are focused behind the galaxy, thus creating a density wake that acts as a potential minimum and exerts tidal forces in the galaxy. In this way, the wake might create asymmetries in both the stellar and gas velocity field and in the distribution of gas and stars in the disc. We test this hypothesis using observations of the isolated galaxy IC 2487 (also called CIG 340). 

\begin{figure*}
    \centering
    \includegraphics{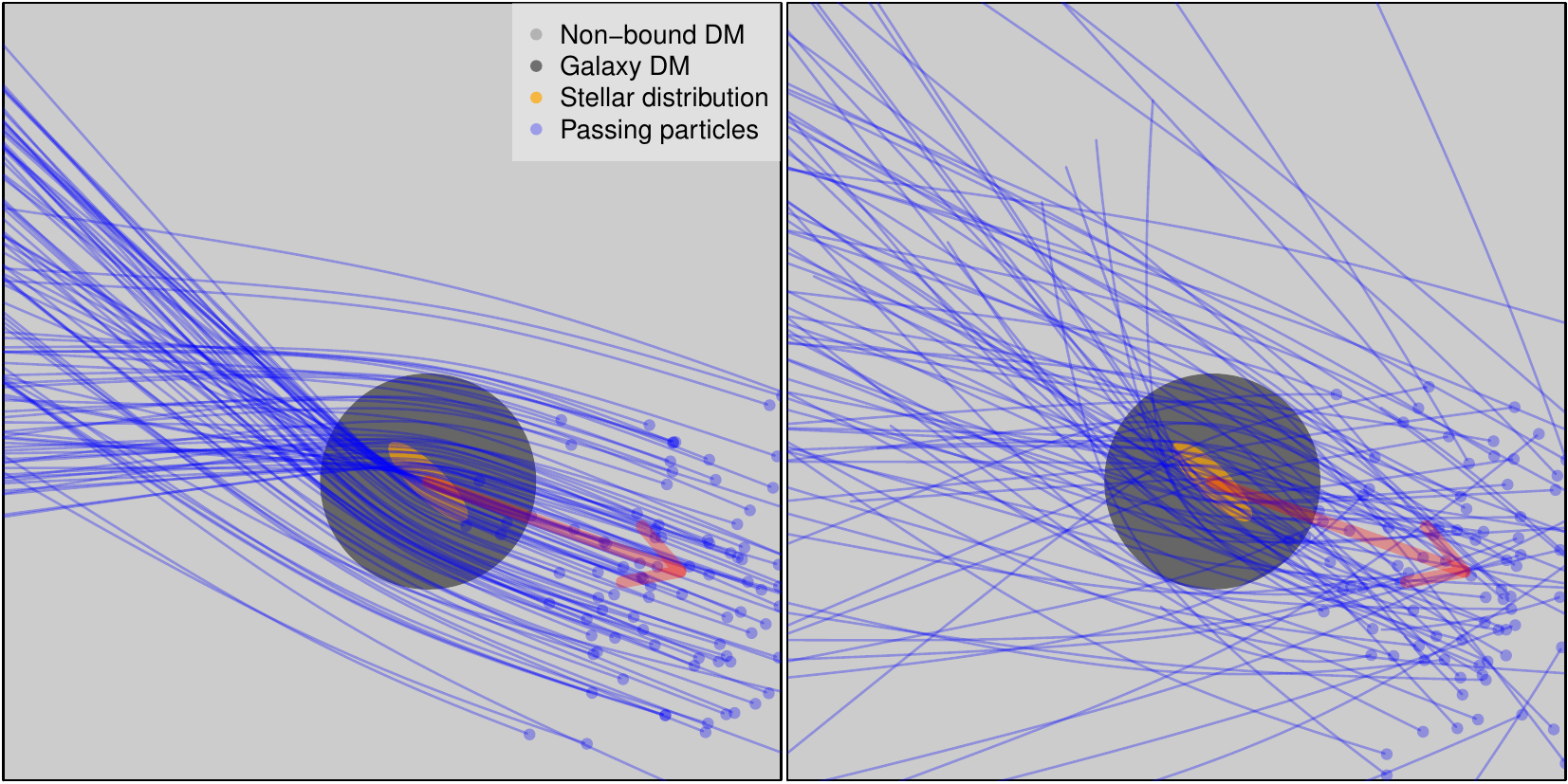}
    \caption{{Illustration of the density wake's production in case of low (left panel) and high (right panel) velocity dispersion of the DM particles.     The trajectory of a sample of DM particles is shown in blue, and the red arrow indicates the direction of motion of the galaxy. 
    }
    }
    \label{fig:schematic}
\end{figure*}

{The properties of the wake depend on the properties of the DM. Fig.~\ref{fig:schematic} schematically shows the formation of the density wake behind the galaxy. The left and right panels differ on the degree of thermal motion (or equivalently the velocity dispersion) of the DM particles. The lower the velocity dispersion, the more concentrated and denser the density wake is. Notice that the wake is quasi-stationary since the passing DM particles do not produce a bound, self-gravitating system. In this work, we do not make any assumption on the origin of the passing DM particles. These might be associated with a particular halo or to the smooth large-scale structure of the Universe.} 

The paper is structured as following: in Sect.~\ref{sec:DF_overview} we review DF and estimate the magnitude of the tidal-like effects and the asymmetries induced in a galaxy. In Sect.~\ref{sec:application} we apply our results to the IC 2487 galaxy (selected as a test-galaxy). Sect.~\ref{sec:discussion} is devoted to the discussion and our paper ends with the summary in Sect.~\ref{sec:summary}.

\section{Dynamical friction}\label{sec:DF_overview}
The dynamical friction (DF) in its classical form is an effective force acting on a  massive point source and being caused by gravitational forces of particles passing by (denoted further as passing particles). Each particle that passes the massive object slightly changes its trajectory due to the exchange of energy and momentum with the massive point source. The cumulative influence on the point source due to all passing particles is called DF \citep{Chandrasekhar:1943} and it acts as an effective force that slows down the massive source.

The derivation of the DF formula assumes that the source and the target of the effect are the same. This means that the point source that causes the change of trajectory of the passing particles is the same as the object for which the acceleration is calculated. Similar results can be calculated when we do not assume this (although such an elegant analytic solution cannot be achieved). In this paper we study the response of test-particles inside a massive object which is the source and target of the effective gravitational deceleration. 

\subsection{Framework and nomenclature}
In the classical formula for DF, the cause of the potential is a point source. In the present study we forfeit this assumption and describe the matter distribution that causes the DF using an extended gravitational potential $\Phi({\bf x})$. The potential is related to the density distribution via the Poisson equation. In the present study we are not interested in the effect of the DF to the overall source of the potential (e.g. galaxy), but to massless test-particles that co-exist in that potential (e.g. stars in the disc of a galaxy). The acceleration of these test-particles depend on their exact position inside the galaxy. Positions, velocities and accelerations of test-particles are denoted as ${\bf x}_\star$, ${\bf v}_\star$, and ${\bf a}_\star$, respectively. For simplicity, we assume that the centre of the reference frame is located at the centre of the potential $\Phi({\bf x})$.

Positions, velocities and accelerations of passing particles (carriers of added momenta) are denoted as ${\bf x}$, ${\bf v}$, and ${\bf a}$, respectively. 
Notice that all the coordinates are time dependent. The mass of these particle is $m$, and their acceleration is calculated from the gravitational potential,
\begin{equation}
    {\bf a}({\bf x}) = -{\bf \nabla}\Phi({\bf x}). \label{eq:galaxy_acc_from_pot}
\end{equation}
Thereafter, their coordinates ${\bf x}$ and velocities ${\bf v}$ are found by calculating their orbits from the equations of motion:
\begin{eqnarray}
    \frac{{\rm d}{\bf x}}{{\rm d}t} &=& {\bf v} \label{eq:eom1}\\
    \frac{{\rm d}{\bf v}}{{\rm d}t} &=& {\bf a}({\bf x}) \label{eq:eom2},
\end{eqnarray}
with initial conditions given by ${\bf x}(t=-\infty) = {\bf x}_0$, $|{\bf x}_0|=\infty$, and ${\bf v}(t=-\infty) = {\bf v}_0$ (where $t$ is time). The second boundary condition means that the passing particles are not bound to the system (i.e. galaxy).

\subsection{Response to a single passing particle}
In this section we estimate how a test-particle, that co-exist in the potential $\Phi({\bf x})$, responds to an event of particle passing. In the present case we assume that the positions of test-particles change only very little per event and can be ignored. Thus, the only effect results from the increase of its velocity. 

A test-particle is accelerated by the gravitational force created by the passing particle. This acceleration can be calculated as 
\begin{equation}
    {\bf a}'_\star({\bf x}_\star, {\bf{x}}) = -\frac{Gm({\bf x}-{\bf x}_\star)}{|{\bf x}-{\bf x}_\star|^3},
\end{equation}
corresponding to the acceleration created by a point source {at time $t$}. Here $G$ is the gravitational constant, and ${\bf x}_\star$ and ${\bf v}_\star$ are the position and velocity of the test-particle, respectively. 

The change in the velocity of the test-particle (i.e. $\Delta{\bf v}$) is calculated by {summing all instantaneous} accelerations during the passing event. Quantitatively, it is found by integrating the force {along} the orbit of the passing particle:
\begin{equation}
    \Delta{\bf v} = \int\limits_{-\infty}^\infty {\bf a}'_\star({\bf x}_\star, {\bf{x}}(t))    {\rm d}t. \label{eq:dv_from_orbit}
\end{equation}
The quantity $\Delta{\bf v}$ describes the response of a test particle to a single passing event. The next step is to include all the passing particles and their relative importance. This is done in the next section. 

\subsection{Cumulative response to passing particles}
Now we calculate the overall effect on a test-particle due to all encounters. The overall acceleration of a test-particle can be found by including all possible events, or equivalently by multiplying the velocity increased per event (i.e. $\Delta{\bf v}$) by the flux of passing particles over an infinitely large surface. That is, 
\begin{equation}
    {\bf a}_\star = \frac{\rho}{m}\int\limits_{-\infty}^\infty f({\bf v}_0) \left[ \int\limits_{A=\infty}
    \Delta {\bf v} ({\bf v}_0\cdot
    {\rm d}{\bf A}) \right]  {\rm d}{\bf v}_0, \label{eq:extra_acc}
\end{equation}
where ${\bf A}$ is the surface of the sphere at infinity (corresponding to the initial conditions for the passing particles), $f({\bf v}_0)$ denotes the initial velocity distribution function of passing particles, with the condition that $\int f({\bf v}_0)\,{\rm d}{\bf v}_0=1$, and $\rho$ is the matter density of the passing particles unperturbed by the potential $\Phi({\bf x})$. {The acceleration field acting on a galaxy -- given in Eq.~\eqref{eq:extra_acc} -- reduces to the Chandrasekhar expression for DF in the case the galaxy is a point source. It is important to notice that our derivation of equation~\eqref{eq:extra_acc} does not require the calculation of the density wake's formation.}

{The acceleration field ${\bf a_\star}$ can be decomposed into three components, i.e 
\begin{equation}
 a_* = a_*^{bulk\,motion} + a_*^{tidal} + a_*^{external}.   
\end{equation}
A first component, $a_*^{bulk\,motion}$, describes the bulk deceleration of the galaxy (which indeed corresponds to the classical DF). A second component corresponds to the tidal field responsible for warp formation and lopsidedness. And finally, a last component accounts for the acceleration field created by matter which is not bound to the galaxy (i.e. the acceleration field created by the passing particles themselves) but is located within the galaxy and contributes to the $M(<r)$ and rotation curve.} This last component might be relevant when, for instance, a galaxy enters a cluster. In such cases, $a_*^{external}$ increases, the resulting acceleration increases, and therefore the radius of the disc truncates {upon entering a cluster. In case of truncation and amount of dark matter halos seen from weak lensing measurements, it is noticed some dependence on the cluster position \citep{Sifon:2018} possibly indicating to a similar mechanism.}

\subsection{Tidal field generated by DM wake}
\label{sec:magnitude_of_effect}

Equation~\eqref{eq:extra_acc} describes the acceleration field exerted on a galaxy which is moving through a background medium of DM particles. It is natural to ask if this field, which is the back reaction of the DM wake, has an important effect on the mass distribution and kinematics of the galaxy. Since the relative perturbations induced by DF to the gravitational potential are smaller than those to the density field \citep{Jog_2009, Binney:2008}, we estimate how large are the corrections to the acceleration field of a galaxy due to DF.

\begin{figure}
    \centering
    \includegraphics[scale=0.5]{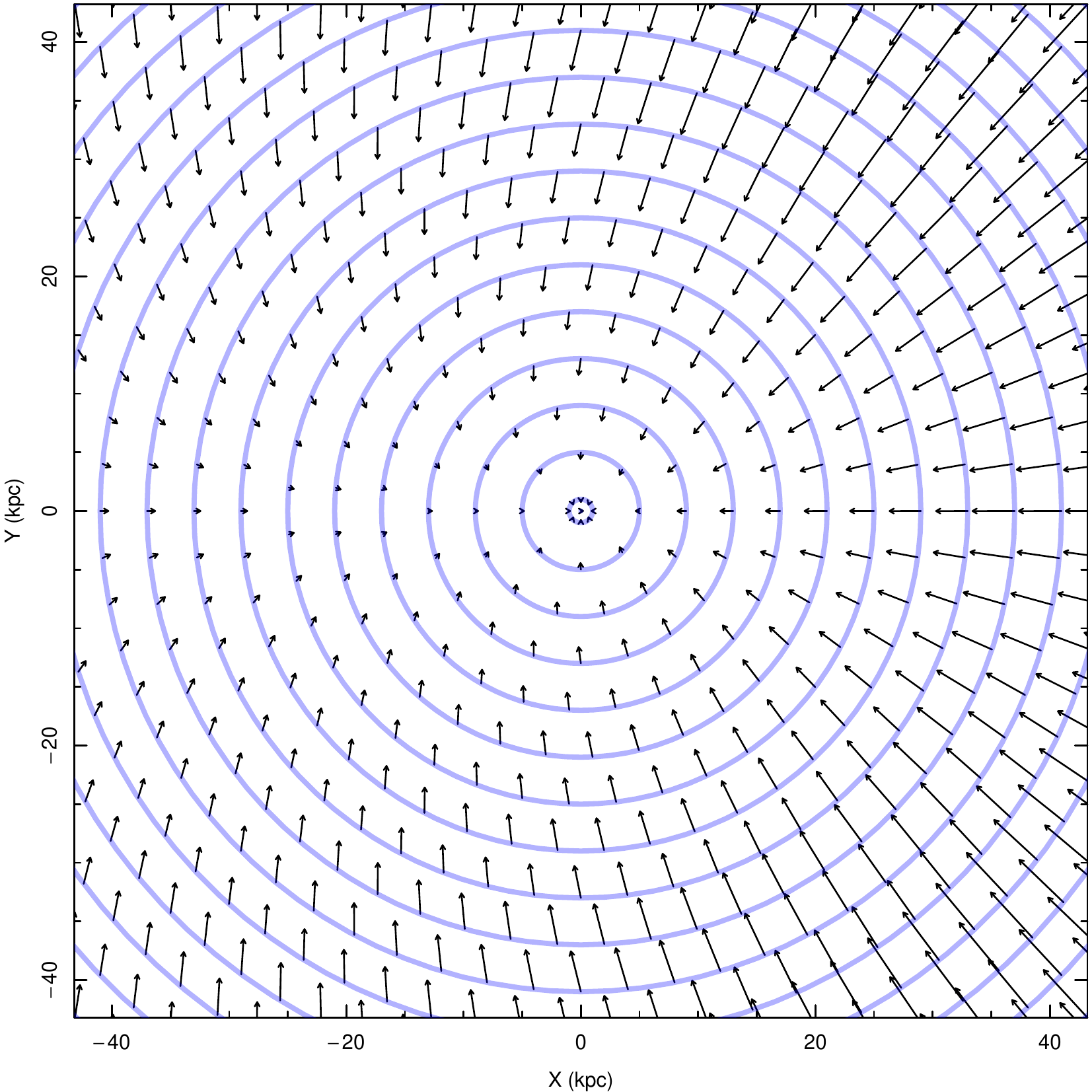}
    \caption{The ${\bf a_\star}$ field in the stellar disc's plane for our test-galaxy with default parameters (see text for details). The arrows show only the direction of the acceleration, the amplitude has arbitrary overall normalisation.
    }
    \label{fig:plane_astar_illustration}
\end{figure}

For this we select a density distribution corresponding to a test-galaxy being described in detail in Sect.~\ref{sec:application}. Our test-galaxy is a disc galaxy with a stellar mass equals to $\log_{10}M_*/{\rm M_\odot} = 10.36$ ($M_*=2.3\times10^{10}\,\rm M_{\odot}$). The magnitude of the tidal field generated by the DM density wake is calculated assuming that the galaxy moves in an environment where the density and the velocity dispersion of the DM particles can be approximated constant along the path of the galaxy for some time. This field depends on four parameters which are estimated to be (onward denoted as default parameters): DM density $\rho_{\rm DM} = 10^5~{\rm M_\odot~kpc^{-3}}$ (average density of the region the galaxy is moving in), isotropic Gaussian velocity distribution of DM particles with velocity dispersion equals to $100~{\rm km\,s^{-1}}$. The speed of the galaxy with respect to the rest frame of DM particles is $v_{\rm gal}=100~{\rm km\,s^{-1}}$  and the angle between the direction of the galaxy's movement and the stellar disc's rotation axis is $\eta = 90^\circ$ (i.e. the galaxy moves edge-on with respect to the background of DM particles). The component of ${\bf a_\star}$ in the stellar disc's plane for the default parameters is shown in Fig.~\ref{fig:plane_astar_illustration}. Notice that the acceleration field is depicted in the galactic reference frame, i.e. field given by equation~\eqref{eq:extra_acc} after subtracting the average acceleration acting on the galaxy. The galactic reference frame is centred at the center of the gravitational potential of the galaxy, $\Phi({\bf x})$. The $x-y$ plane coincides with the stellar disc's plane and the $z$-axis equals the stellar disc's rotation axis. The galaxy is moving towards positive x-axis. 

\begin{figure*}
    \centering
    \includegraphics{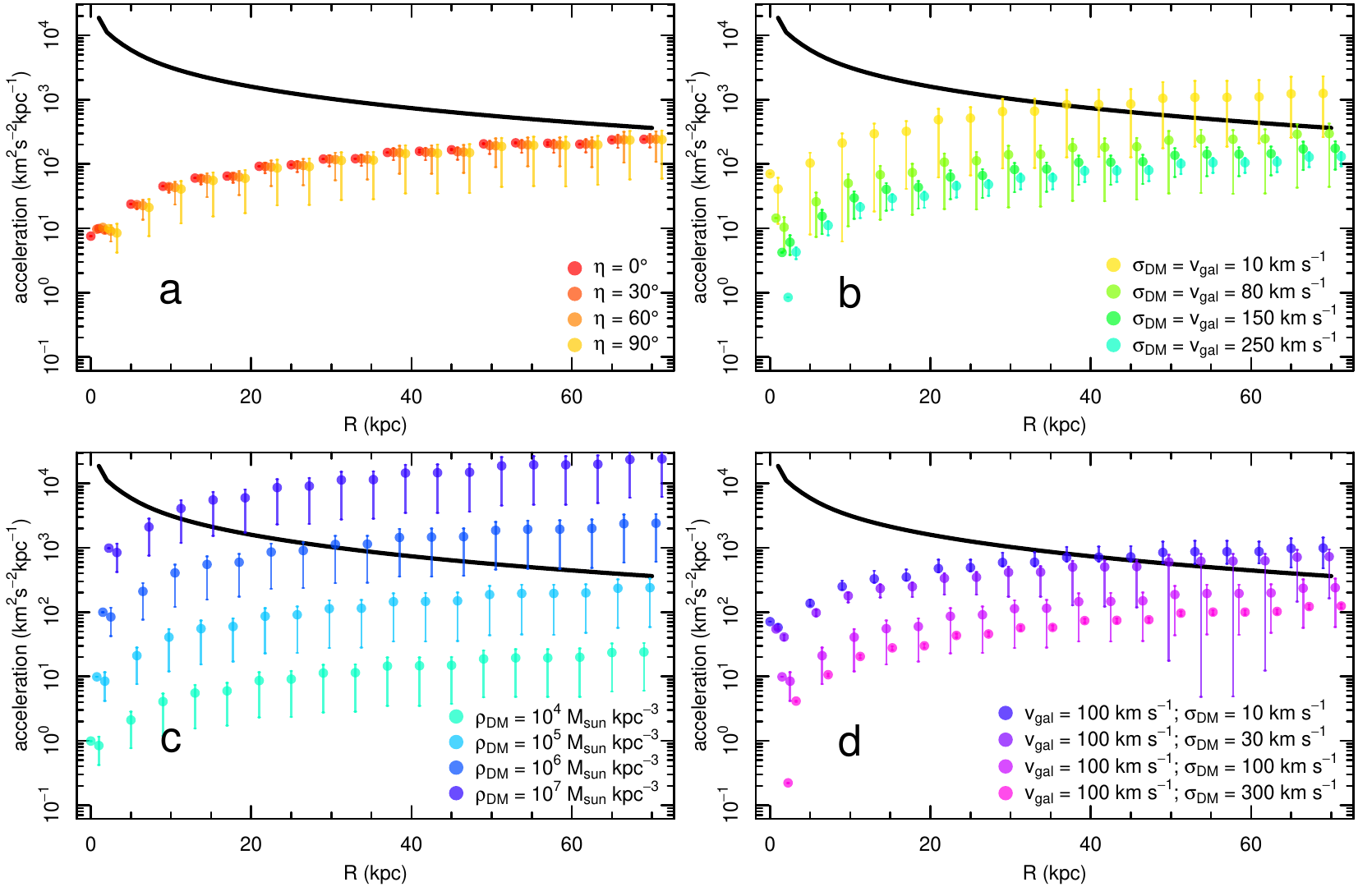}
    \caption{These figures illustrate dependencies of the magnitude of $a_\star$'s component in the disc's plane with respect to different parameters. Black lines represent the galaxy's acceleration given by Eq.~\eqref{eq:galaxy_acc_from_pot}, colorful points correspond to the averaged value of $|{\bf a_\star}|$ in the galactic reference frame at constant radii and {the bars encompass the values this parameter takes at each radii. Indeed, it is this range of values or variation which decides the resulted asymmetry.} The {panel a} shows the dependence on the angle between the movement of the galaxy and the stellar disc's rotation axis (with $\eta = 90^\circ$ meaning that the galaxy moves edge-on with respect to the DM halo). The {panel b} shows how $|{\bf a_\star}|$ changes with varying  galaxy's velocity and velocity dispersion of DM particles, but with fixed ratio between these two parameters. The f{panel c} shows the DM density dependence, and in the {panel d} the dependency with the DM velocity dispersion is shown. For visualization purposes, the points are slightly shifted in the x-axis. 
    }
    \label{fig:all_dep_astar}
\end{figure*}

We vary each of the above parameters individually while keeping the others fix to their default values. The specific dependencies on galaxy and environmental properties of the ${\bf a_\star}$ component in the disc's plane  are shown in Fig.~\ref{fig:all_dep_astar} as a function of galactocentric distance $R$. 
From the upper left panel, we can see that the sizes of the {bars (which indicate the spread in ${\bf a_\star}$ values at given radii)} of $|{\bf a_\star}|$
\footnote{Notice that by $|{\bf a_\star}|$ we denote the magnitude of the component of $a_\star$ in the disc's plane and not the magnitude of the 3-dimensional $a_\star$.} change with the angle between the direction of movement of the galaxy and the stellar disc's rotation axis. This results from the relative contribution of different vector components. For instance, if the galaxy's speed and the disc are perpendicular (i.e. $\eta=0$, or face-on movement), all the points in the galactic plane at a given radius are at the same distance from the wake, thus there is axial symmetry (i.e. there are no bars). Furthermore, in this case the outer parts of the disc are pushed upward (in the disc reference frame) with respect to the inner parts. This might result in U-shaped warps, contrary to S-shaped warps that can be produced when moving with larger $\eta$ (more edge-on with respect to DM). { Indeed, with larger $\eta$ values, the amplitude of the asymmetry might be different between the two sides of the disc, with the side closer to the amplitude showing a higher amplitude. Thus, L-shaped warps are more likely in this mechanism. Notice that galaxies are known to show asymmetry in the warps (e.g. \cite{Saha:2006}), whose origin is not well-understood. } In the case of edge-on movement (i.e. $\eta=90^\circ$), the range of values that $|{\bf a_\star}|$ takes at constant $R$ is larger. As commonly in the case of tidal forces, each galaxy's side is accelerated to different directions {and, thus, more values are possible.} {In a subsequent analysis, we will demonstrate that depending on the strength of the disc's response, some fraction of warps can show both \textit{S} and \textit{U}-shaped behaviour, depending on radius.}

From the lower left panel of Fig.~\ref{fig:all_dep_astar}, we see that $\rho_{\rm DM}$ acts as a normalization of the tidal field exerted on the galaxy. This can also be seen from equation~\eqref{eq:extra_acc}. 

From the lower right panel of Fig.~\ref{fig:all_dep_astar}, it is seen that the acceleration ${\bf a_\star}$ is sensitive to the velocity dispersion of DM particles. In most cases the colder the dark matter particles produce  larger ${\bf a_\star}$ values, although not always guaranteed to be so. The departure from axial symmetry of ${\bf a_\star}$ strongly depends on the relative kinematics between the galaxy's movement and the velocity dispersion of DM particles. From the bottom right panel we can see that the colder DM suggests higher ${\bf a_\star}$ values, but does not {resolve} the question, whether the effect originates from velocity dispersion or from the galaxy velocity (i.e. Chandrasekhar formula for DF in isotropic case has no independent velocity dispersion term, but only via ratio of galaxy movement and velocity dispersion). The top-right panel of the figure show the effect of the DM velocity dispersion when it is tied with the galaxy velocity. The trends show increase of DF when velocities are decreased.


\subsection{Response of a galaxy to the tidal field}

DF acts on systems in very different environments, such as globular clusters in halos or the planetesimal movement in protoplanetary discs. In the present section we concentrate on DF acting on galaxies moving in a field of DM particles. In particular, we estimate the back response of their stellar disc to the tidal field induced by the DM wake which is responsible for DF. 

The tidal field generated by the wake of DM particles modifies the orbits of the stars in the galaxy (see Eq.~\eqref{eq:extra_acc} and Section~\ref{sec:magnitude_of_effect}). In most of the cases, ${\bf a}_\star$ is non-axisymmetric, therefore the response of the disc might not be trivial to calculate. To estimate the modified outlook of the galaxy, we assume that disc's stars initially move in circular orbits and we calculate how these orbits change due to the additional ${\bf a_\star}$ field given by Eq.~\eqref{eq:extra_acc}.

We expect the tidal effects to be strongest in the outer parts of the disc. Since the best tracer of the outer disc is gas, we aim to build up our further modelling by taking the gas component into account. The collisional nature of the gas supports its settlement into a thin disc rotating approximately according to circular orbits. When calculating the response of the galaxy to tidal effects, we assume that initial conditions correspond to circular motions with $v_{\rm circ}=\sqrt{a\cdot R}$. Subsequent evolution of disc particles is described with the Eq.~\eqref{eq:eom1} and 
\begin{equation}
    \frac{{\rm d}{\bf v}}{{\rm d}t} = {\bf a}({\bf x}) + {\bf a}_\star({\bf x}).
    \label{eq:eom3}
\end{equation}

In any orbit, each orbital segment is observable with a probability proportional to the time an object in this orbit spends in the corresponding segment \citep{Han:2016}. This allows to construct the change of the orbit from the initial circular state to the modified state. Therefore we are able to see the response of the galactic disc to the tidal field. 

\begin{figure*}
    \centering
    \includegraphics{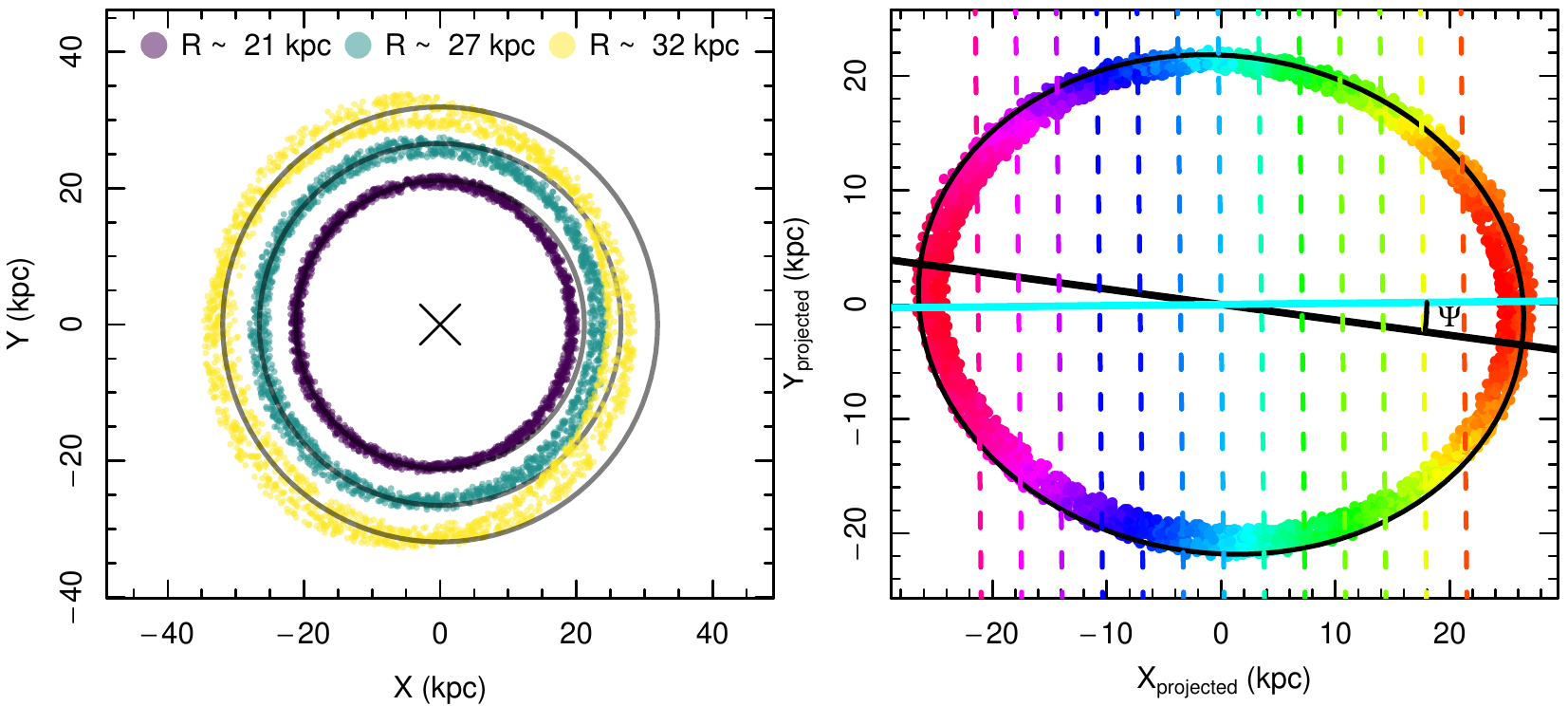}
    \caption{Left panel: three orbital families (i.e. many orbits with slightly different initial conditions) modified by tidal effects. Grey circles correspond to the initial circular orbits with initial radii indicated at the top of the figure. Different responsiveness to the $a_\star$ catches the eye. Right panel: an orbital family observed at an angle of $35^\circ$ with misalignment to the photometric axis of $\Psi = 8^\circ$. Cyan line is the estimated kinematic axis, black line is the photometric (initial) axis. The colored points show line-of-sight velocity of each point. Dashed lines connect similar velocity regions, and help to assess their orientation in the plane of sky. This illustrates {ease} of creation of the position angle misalignment via tidal effects. 
    }
    \label{fig:orbit_observables}
\end{figure*}

An example of the orbital families\footnote{In this work, an orbital family refers to the superposition of many orbits with slightly different initial conditions.} used in this work is presented in Fig.~\ref{fig:orbit_observables}. For orbit calculations we use the default parameters from Sect.~\ref{sec:magnitude_of_effect}. In the left-hand panel of this figure it is seen that the outermost orbits are more strongly distorted to lopsided shapes, which is a common observational feature of many galaxies. The right-hand panel illustrates the misalignment of kinematic and photometric major axes of the galaxy ($\Psi$). Here the orbit is projected at $35^\circ$ to bring out the line-of-sight velocity components. The misalignment results in a natural way up to a values where there are difficulties in determining the position angles. 

\section{Application to galaxy IC 2487}
\label{sec:application}
{We quantify the effects that the tidal field, which is generated by the density wake of DF, have on the galaxy IC 2487 (also known as CIG 340). 
This galaxy} is a disc galaxy located at a distance of $54.8$~Mpc \citep{FL:2013} and is either an isolated or lies in a sparse group. We select an isolated galaxy\footnote{The nearest neighbour to the IC 2487 galaxy is at $0.6$ Mpc in projection, which (in combination with typical line-of-sight velocities) is equivalent to $2.9$ Gyr of motion \citep{Scott:2014}.} in order to reduce other possible phenomena responsible for lopsidedness or warp production. Hence increasing the probability that {the density wake of DF} is the cause of these asymmetries. The IC 2487 galaxy is well observed. It is covered by SDSS imaging \citep{sdss_dr12}, the CALIFA \citep{califa_dr2} and the AMIGA\footnote{AMIGA project homepage is \url{http://amiga.iaa.es/p/1-homepage.htm}} \citep{amiga:2005} surveys. \cite{Scott:2014} studied the HI distribution of this galaxy, emphasising the asymmetry of its gas distribution. 

Throughout this Section we denote $r$ as a spherical coordinate, and $R$ and $z$ as cylindrical coordinates. The centres of both spherical and cylindrical reference frames are co-aligned with the centre of the IC 2487 galaxy. 

For quantifying the effect of tidal forces in the galaxy, we, first, model the ability of IC 2487 to generate a DM density wake by gravitational scattering of DM particles. The generation of this wake depends on the gravitational potential of the galaxy $\Phi({\bf x})$ which can be decomposed into a stellar and a DM components. Secondly, we model the lopsidedness in the HI distribution and velocity field of the IC 2487 galaxy using measurements of the 21-cm line emission.

\subsection{Gravitational potential of IC 2487}
\label{subsec:IC2487_phi}
\begin{figure}
    \centering
    \includegraphics{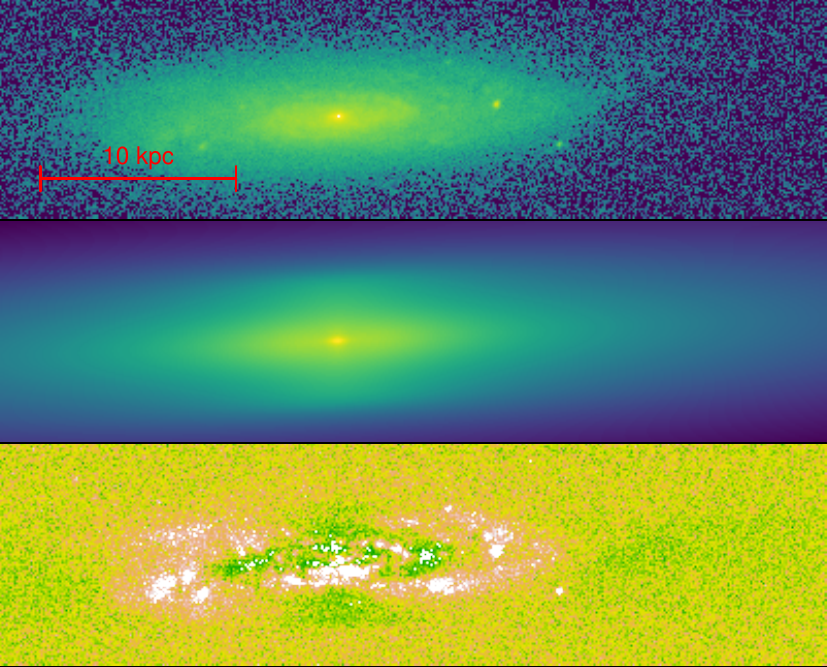}
    \caption{
    {Stellar distribution of the IC 2487 galaxy.}
    The top panel shows the optical image observed by the SDSS survey in the $i$ band, while the middle panel corresponds to our model image. {In both cases the color scale is logarithmic. The bottom panel shows the residuals in linear scale.}
    }
    \label{fig:ic2487_sdss}
\end{figure}

The matter distribution of a galaxy can be decomposed into three components: gas, stellar mass and DM. For simplicity, and since the contribution of the gas to the total gravitational potential is in the present case rather small, we do not take it into account in the mass modelling. The stellar component is modelled using the SDSS $i$-band image which can be seen in the top panel of Fig.~\ref{fig:ic2487_sdss}. 

The stellar mass distribution is described by two co-centric Miyamoto-Nagai profiles \citep{MN1975}, one describing the bulge and the other describing the stellar disc component. Although an exponential density profile better fits the luminosity distribution of stellar discs  in galaxies, it lacks an analytical form for both acceleration and gravitational potential. Thus, we select a Miyamoto-Nagai profile as it has an analytical potential form given by 
\begin{equation}
    \Phi_{\rm MN}(R, z) = \frac{-GM_{\rm MN}}{\sqrt{R^2 + \left(a + \sqrt{b^2 + z^2}\right)^2 }}  \label{eq:MN_pot},
\end{equation}
with the corresponding spatial mass density tie with the potential via the Poisson equation. For the mass modelling, we include the line-of-sight integration, but not the intrinsic dust absorption, which we consider insignificant for the $i$-filter. The model image obtained by photometric modelling is given in the middle panel of Fig.~\ref{fig:ic2487_sdss}. The lower panel shows the residual image of the observations and the model. We found that the IC 2487 galaxy is best described by a disc component with $a = 2.2\,{\rm kpc}$ and $b = 0.6\,{\rm kpc}$, and a bulge component with $a = 0.2\,{\rm kpc}$ and $b = 0.1\,{\rm kpc}$. {The bulge-to-total stellar mass fraction, which is assumed to match the luminosity fraction, is 0.12. The total stellar mass is taken to be $\log_{10}M_*/M_\odot = 10.36$ ($M_*=2.3\times10^{10}\,\rm M_{\odot}$) \citep{FL:2013}. } The {resulting} inclination angle is $83.9^\circ$, which is in good agreement with the value of $83.7^\circ$ found by \cite{Scott:2014}.

The DM density distribution of the IC 2487 galaxy is modelled by fitting the major axis stellar rotation plateau measured by the CALIFA survey. As it is based on stellar kinematics, we increased rotation velocities by a typical value of $\sim10\,{\rm km\,s^{-1}}$ to include asymmetric drift and line-of-sight integration effects. The value of $10\,{\rm km\,s^{-1}}$ was chosen in order to tie CALIFA and HI rotation curves. The DM distribution is described by a NFW profile \citep{NFW1996} with density and potential profiles given by
\begin{eqnarray}
\rho_{\rm NFW}(r) &=& \frac{ \rho_s}{  \frac{r}{r_s}\left( 1 + \frac{r}{r_s} \right)^2 }  \\
\Phi_{\rm NFW}(r) &=& \frac{4\pi G \rho_s r_s^3}{r}\ln{\left(1 + \frac{r}{r_s}\right)},
\end{eqnarray}
where $\rho_s$ and $r_s$ are the scale density and scale radius, respectively. 
{As found in DM-only N-body simulations, DM density profiles towards the centre of halos become more gradually shallower than predicted by the NFW profile. In order to account for this effect, an Einasto profile \citep{1965TrAlm...5...87E} was proposed to describe the mass density of simulated halos (e.g. \cite{Merritt:2006, Gao:2008}). Nonetheless, in this work we use a NFW profile since it has an analytic form for its corresponding gravitational potential.} The resulting DM density profile has $\rho_s = 1.4\times10^7\,{\rm M_\odot\,kpc^{-3}}$($=\SI{0.53}{GeV/cm^3}$) and $r_s = 13\,{\rm kpc}$. 

Finally, the total gravitational potential of the galaxy is the sum of all the stellar and DM components,
\begin{equation}
    \Phi = \Phi_{\rm NFW} + \sum_i\Phi_{i,{\rm MN}},\label{eq:pot_ic2487}
\end{equation}
where $i$ indexes the bulge and disc components.

\subsection{Observations}
In Section \ref{sec:magnitude_of_effect} we found that the tidal acceleration field induced by the DM wake responsible for DF is larger in the outer parts of galaxies (e.g. see Fig.~\ref{fig:all_dep_astar}). Since the response of the galaxy to tidal effects is, thus, prone to be larger in the outer regions, it is best to look for data that extends to large radii. Radio observations of the HI distribution in galaxies fulfil this criterion. 

Observations of the 21-cm HI emission line for the IC 2487 galaxy are taken from \citet{Scott:2014}. In particular, we use the flux density in a collection of three-dimensional pixels for two different resolution regimes (i.e. low and high resolution). Each pixel is characterised by the two-dimensional position and line-of-sight velocity. Thus, the tidal field, induced by the DM wake responsible for DF, acting on the IC 2487 galaxy is determined by modelling the observed flux as a function of the two-dimensional position and the line-of-sight velocity, i.e. $I(X, Y, v_{\rm los})$.

A complementary part of the observations is the point-spread function (PSF) of the imaging. We approximate it with the ${\rm sinc}^2$ function, with full width at half maximum (FWHM) taken from \citet{Scott:2014}. We assume that the PSF is a top-hat function in velocity space.
Uncertainties on the flux density are found from empty background and take the following values: $\sigma_1 = 1.4$~mJy and $\sigma_2 = 1.0$~mJy for the low and high resolution regimes, respectively. We add a slight uncertainty proportional to the flux density in each pixel $\sqrt{\sigma^2 + 2\sigma|I|}$, where $\sigma$ and $I$ are, respectively, the uncertainty and flux of each pixel. Since the uncertainties are correlated pixel by pixel, we increase the amplitude of the uncertainty by {a factor of 2 in order} to avoid over-fitting. A detailed description of the data can be found in \citet{Scott:2014}.

\subsection{Model and statistical framework}
\label{sec:LLstability}
We determine the tidal field by modelling the flux density as a function of the two-dimensional position and the line-of-sight velocity. The degree of similarity between observed and model fluxes is measured by means of the likelihood function 
\begin{equation}
    \log\mathcal{L} = -\sum_i\left[ \frac{(I_i-I_{i,{\rm mdl}})^2}{2\sigma_i^2} \right],\label{eq:LL}
\end{equation}
where the summation is done over all three-dimensional pixels. The model flux at each pixel $I_{i, \rm mdl}$ is given by the convolution of the superposition of different orbital families with the PSF. 
We include $15$ orbital families covering initial radii from $2$ to $45$ kpc. 
Each orbital family consists of $500$ orbits with similar initial conditions. Each orbit is determined by $100$ points taken uniformly in time. Thus, each orbital family is represented by $5\times10^4$ points. We assume initial circular orbits in {the gravitational potential estimated in Sec.~\ref{subsec:IC2487_phi}.\footnote{
{The assumption that the non-perturbed potential equals the one determined today has a two-fold justification. Firstly, the depth of the potential is not affected by the tidal force. Secondly, today's potential is determined from measurements of the inner part of the galaxy where tidal effects are not important.}} These orbits have a radius and initial angle determined by the initial conditions of the orbital family which are characterised by a radius range with a width similar to the PSF's radius.} The radii and phase angles of initial orbits are taken randomly from uniform distribution. 

The initial system of orbital families is evolved in time using a Dormand-Prince algorithm. To mimic the build-up of the DM wake, we assume that in the initial $t=0.25$~Gyr, the total acceleration
acting on the galaxy is ${\bf a} + {\bf a_\star} t/[0.25~{\rm Gyr}]$, where ${\bf a}$ is the acceleration generated by the gravitational potential of the galaxy (i.e. equation~\eqref{eq:galaxy_acc_from_pot}) and ${\bf a_\star}$ is given by equation~\eqref{eq:extra_acc}. After the first 0.25 Gyr, the acceleration field acting on the galaxy is ${\bf a} + {\bf a_\star}$. The outlook of the orbits/galaxy are dependent on the timescale the perturbation has occurred. As we do not know it from observations, the model flux is computed after a time interval of 1.25 and 2.25 Gyr. We select two different orbital integration times to test the sensitivity of our results to this parameter.

Once the system is evolved in time, the model flux is given the superposition of projected orbital families that is convolved with the PSF:
\begin{equation}
    I_{i,{\rm mdl}} = \sum_k P_{i,k} * \sum_j w_j O_{j,k},
\end{equation}
where $i$ and $j$ run over pixel and orbital family, respectively, and $k$ represents integration of the PSF. Furthermore, $P$ and $*$ denote PSF and convolution, $w_j$ is the weight of each orbital family, and $O$ denotes the set of points that represents each orbital family. The model flux is a function of the following six free parameters: the density and velocity dispersion of background DM particles given by $\rho_{\rm DM}$ and $\sigma_{\rm DM}$, respectively. We assume a Gaussian velocity distribution of DM particles. The velocity of the galaxy with respect to the background medium which is characterized by $v_{\rm gal}$ and the angles $\eta$, $\phi$. Finally, the galaxy's redshift is also a free parameter of the analysis. For each likelihood evaluation, the weights $w_j$ are found by minimising the difference of $I$ and $I_{\rm mdl}$ by means of a Newton-Rhapson method which keeps the orbital weights positive. 

The modelling of tidal effects implies the calculation of a five-dimensional integral and orbit integration both for Eq.~\eqref{eq:dv_from_orbit} and for the later response of the galaxy, seven nested integrations in total. Each subsequent integration produces numerical inaccuracies, and therefore the evaluation of the likelihood has numerical inaccuracies. We caution that the exact numerical values should not be taken as very precise ones, but with slight caution.

\subsection{Results and inference}

\subsubsection{Quality of the model}

\begin{figure*}
    \centering
    \includegraphics{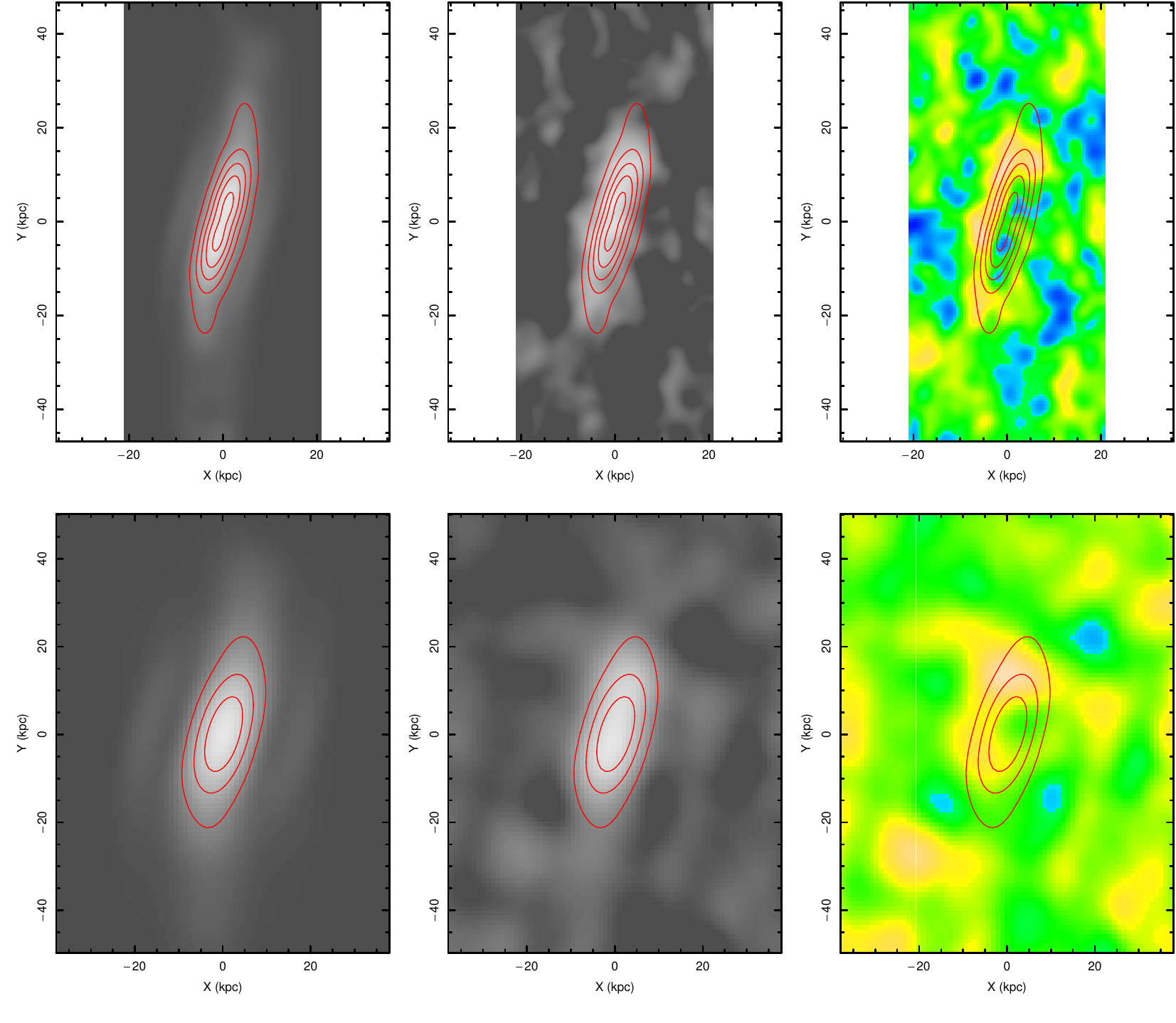}
    \caption{
    {HI emission maps (i.e. flux integrated over line-of-sight velocities). Top and bottom rows corresponds to the high and low resolution maps, respectively.} Left panels are model images, central ones are observed images, and the right ones correspond to the residuals. The red contours are taken from the model images for easier comparison. The model is obtained after an integration time of $2.25$ Gyr (see text for details).}
    \label{fig:ic2487_mdl_image}
\end{figure*}

The technique for modelling the HI distribution of the IC~2487 galaxy was presented in the previous sections. Although the modelling has been done by fitting the observed flux as a function of two-dimensional position and line-of-sight velocity, we are only able to show two-dimensional projections for a reader to assess the quality of the fit. The projection into the sky's plane {(i.e. flux density integrated over velocity)} is shown in Fig.~\ref{fig:ic2487_mdl_image}.
From this figure it is seen that, for both high and low resolution images, the modelled warp comes forth. The model images for the low and high resolution regimes are slightly different since the PSFs for them are differently oriented: $15.5^\circ$ and  $-59.73^\circ$ \citep{Scott:2014}. 

\begin{figure*}
\centering
\includegraphics{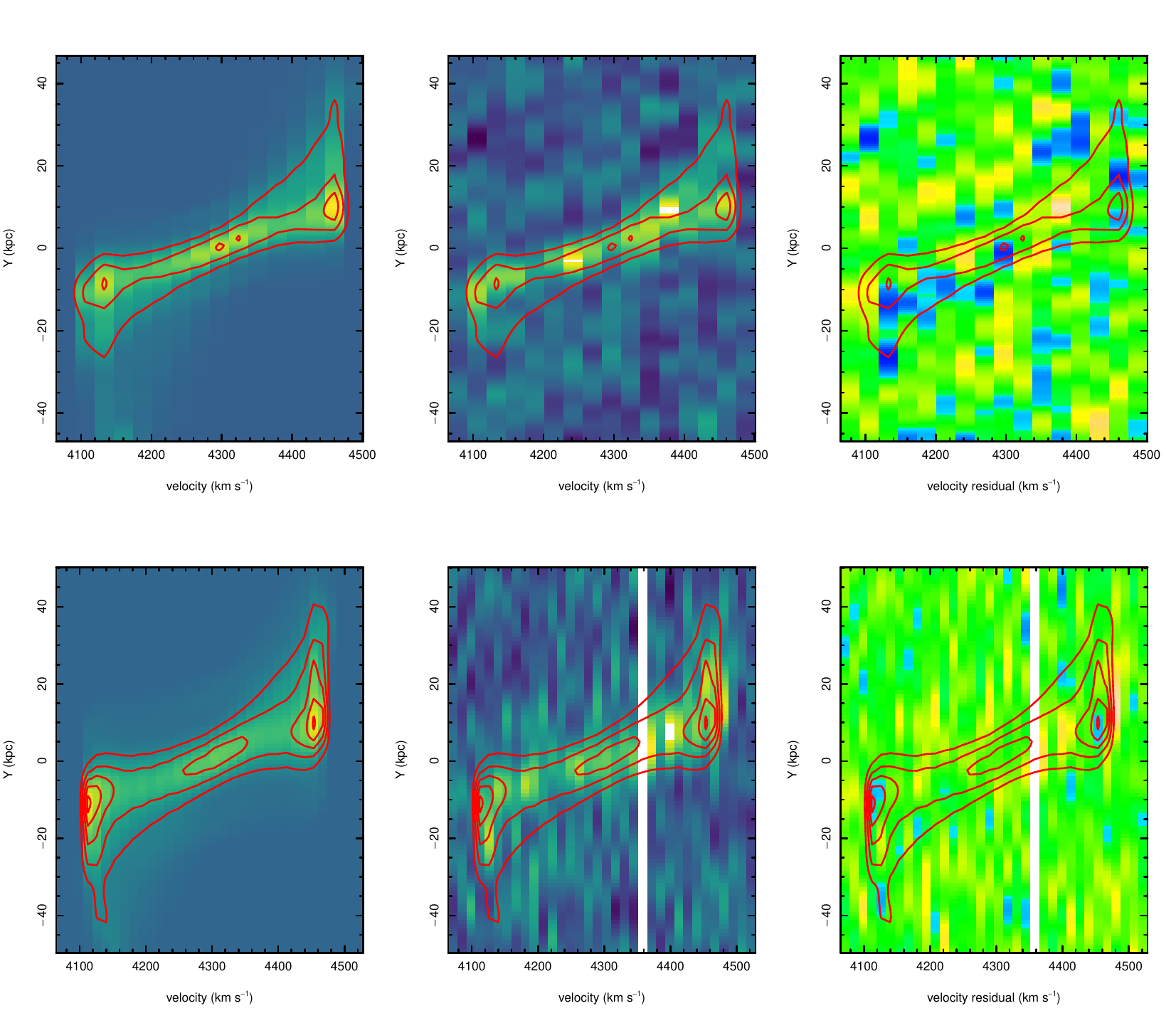}
\caption{
{HI velocity field obtained by summing over the spatial $X$-direction.}
Top panels show the high resolution images and bottom ones show the low resolution (but better resolved in velocity space). From left to right are model, observational and residual images, where the contours from the model image have been added. 
}\label{fig:ic2487_mdl_vel}
\end{figure*}

The velocity distribution model can be assessed from Fig.~\ref{fig:ic2487_mdl_vel}. In this figure we show the projection of spatial velocities summed over the $X$-direction in the HI observation data-cubes, leaving the velocity dependence over the $Y$-direction (vertical direction in Fig.~\ref{fig:ic2487_mdl_image}). We conclude that model images match observations well for both low and high resolution regimes. 

\begin{figure*}
    \centering
    \includegraphics{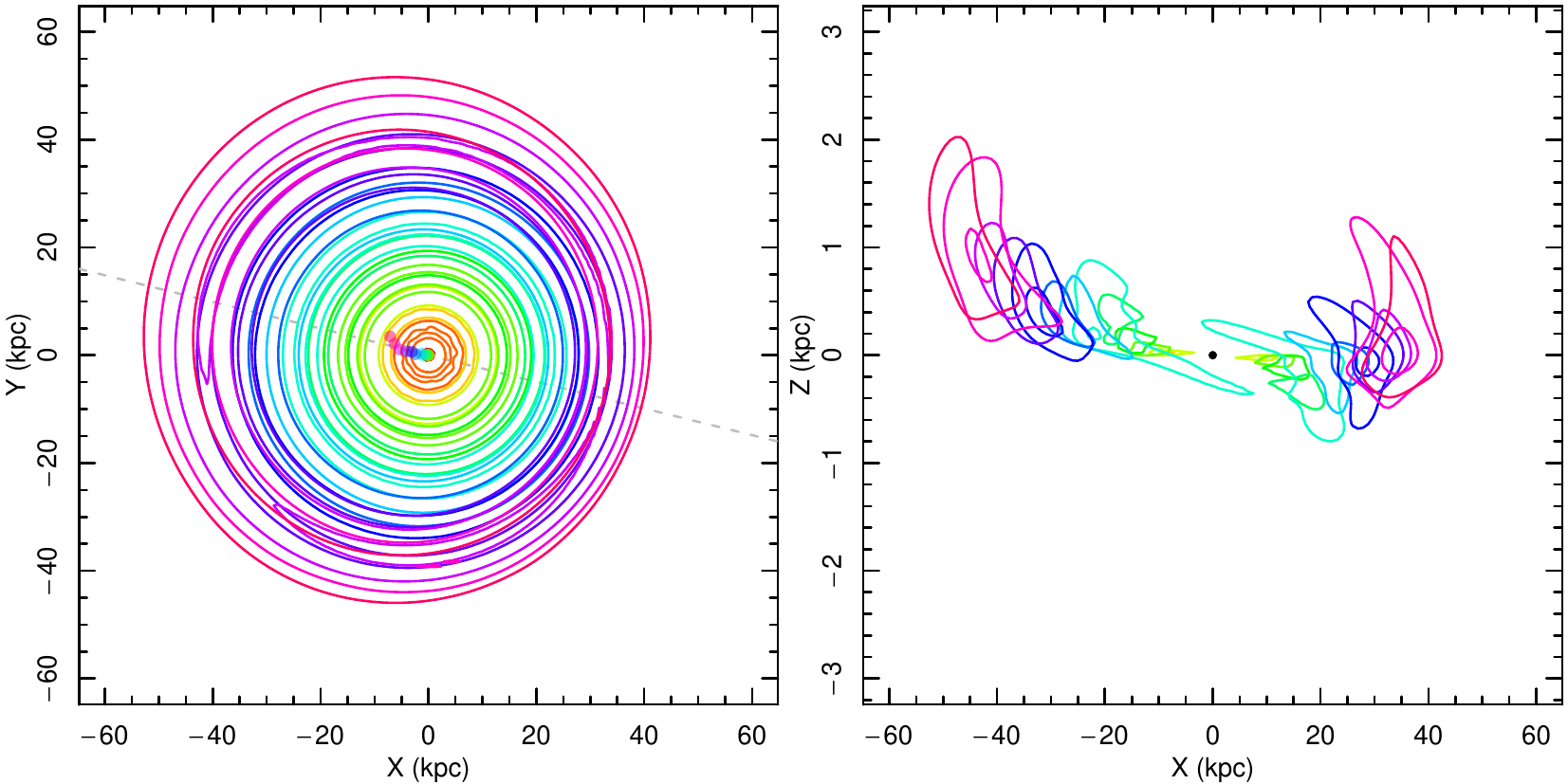}
    \caption{{Superposition of orbital families contributing to IC 2487. The left panel shows the face-on view and the lopsidedness of the galaxy. The dots near the centre show the central values of each orbital family. As they are aligned, there is no significant phase shift. The right hand panel shows the edge-on projection. The inner parts show practically no warp, the middle ones \textit{S}-shaped one, and outer regions $U$-shaped warp. }}
    \label{fig:ic2487_family}
\end{figure*}

\begin{figure}
    \centering
    \includegraphics{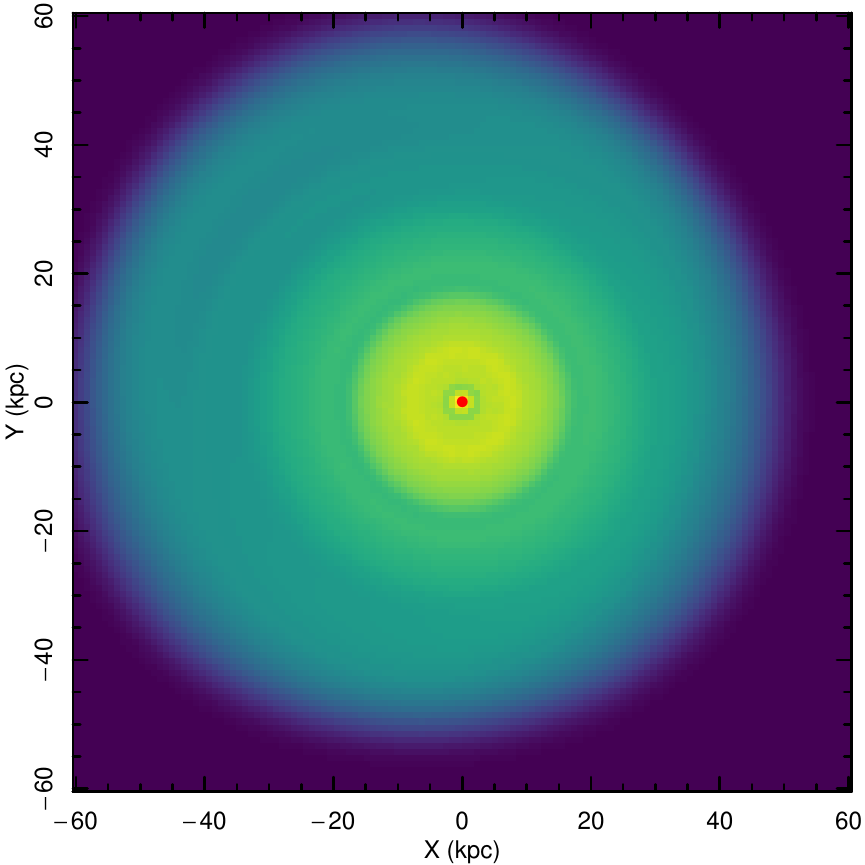}
    \caption{{Best model for IC 2487 in face-on projection without PSF convolution. It is noticeable that the centre is quite round while outer parts show significant lopsidedness.}}
    \label{fig:faceon_best_model}
\end{figure}

{We saw that the model resembles the observations very well. An advantage of the model compared to observations is that we can project it in different ways to emphasize different aspects. Firstly we would like to show the orbital families that build up the gas distribution. This is shown in Fig.~\ref{fig:ic2487_family}. The left panel shows the face-on view of the galaxy. The colorful dots show the centres of each of the orbital families, their alignment illustrates no significant phase shift with radius. The panel on the right shows warpness. At central regions, no warp is seen, at intermediate radii, there is \textit{S}-shaped warp, and at outermost radii, {the orbital families form an} \textit{U}-shaped warp. This is caused since the vertical acceleration of the disc is insignificant compared to the force produced from the wake. Although interesting shift from \textit{S}-shaped warp to \textit{U}-shaped warp, it might be not well observed because of the high distance from the centre. In Fig.~\ref{fig:ic2487_family} we depicted only orbital families, ignoring their relative strength. In Fig.~\ref{fig:faceon_best_model} we have adjusted the orbital family weights and provide the face-on projection in logarithmic intensity scale. The lopsidedness is well shown in this figure.}

\begin{figure}
    \centering
    \includegraphics{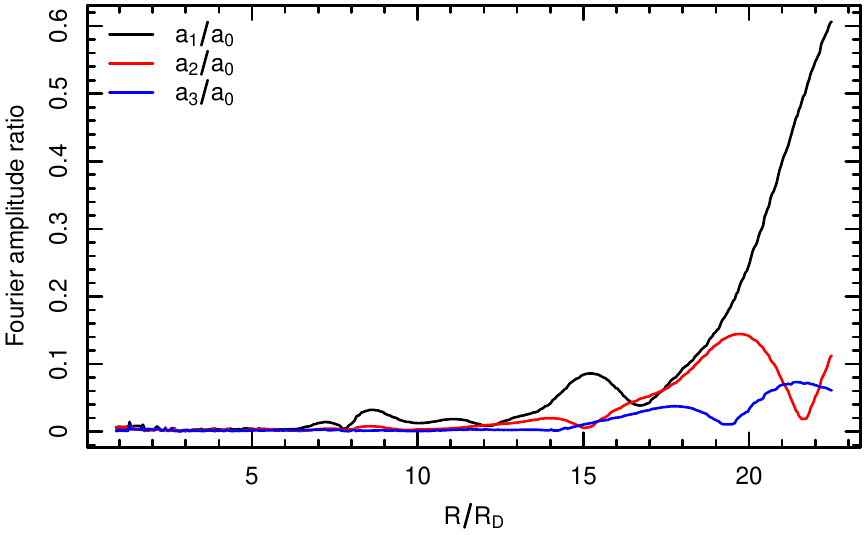}
    \caption{{Fourier amplitudes for model IC 2487. {Amplitudes are spiky} because the orbits are not smoothed by PSF/beam. }}
    \label{fig:fourier_amplitudes}
\end{figure}

{The observationally motivated measures of asymmetry are ratios of coefficients from Fourier series. We projected our model to a face-on view, and expanded the intensities over constant radii circles into Fourier series. The ratio of the coefficients quantify each type of asymmetry: $a_1/a_0$ lopsidedness, and $a_2/a_0$ flatness or the bar-likeness. The measured asymmetry can be seen in Fig.~\ref{fig:fourier_amplitudes}. We would like to point out that the gas distribution has a shallow gradient, indicating that the orbital asymmetry (i.e. non-roundness of lines in the left panel of  Fig.~\ref{fig:faceon_best_model}) can be larger than the intensity-based measures.} {A comparison between the theoretical lopsidedness as shown in Fig.~\ref{fig:fourier_amplitudes} and observations from \citet{2011A&A...530A..30V} show different behaviour at very large radii: whether in the case of observations, the $a_1/a_0$ fraction saturates, in our model it does not. A possible cause for this is that our wake calculations do not include self-gravity of the wake and possible lopsidedness of the potential from the response of the galaxy (see Sect.~\ref{sec:caveats}).}

\begin{figure}
    \centering
    \includegraphics{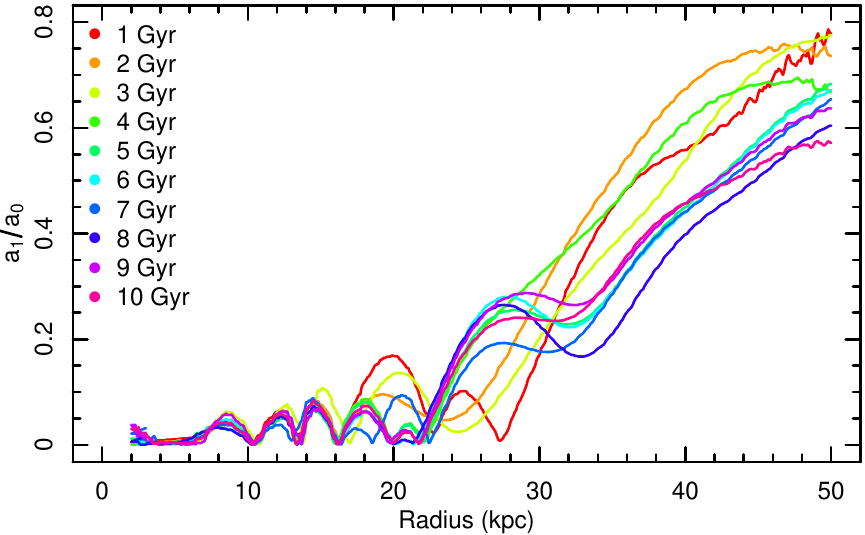}
    \caption{{This figure shows the build-up of lopsidedness over $10$~Gyr. The response of the disc stabilises to a quasi-stationary state above $\sim 6$ Gyr. In case of strong symmetries of potential, the orbits tend to be more stable, but in the present case we are at the opposite end: potential does not follow even tri-axial symmetry. 
    }}
    \label{fig:stability_over_integration_time}
\end{figure}

{The integration time was one of the parameters of modelling. We selected it to be 1 and 2 Gyr since this is the time-scale that a galaxy would take to travel through a cluster or group of galaxies. Thus, within this time we can assume that the phase-space distribution of the passing particles is constant.
We assessed how the asymmetry changes when the disc responds for $1 - 10~{\rm Gyr}$ and show it in Fig.~\ref{fig:stability_over_integration_time}. We can see that initially there are some changes, but after $\sim6$~Gyr, the system seems to stabilise.}

\subsubsection{Inference on galaxy movement and dark matter}
Our modelling of the HI distribution in the IC 2487 galaxy has six free parameters: the galaxy's receding velocity from us (i.e. the redshift), its velocity with respect to the DM background (three vector components), the DM density and the DM velocity dispersion. 

\begin{figure}
    \centering
    \includegraphics{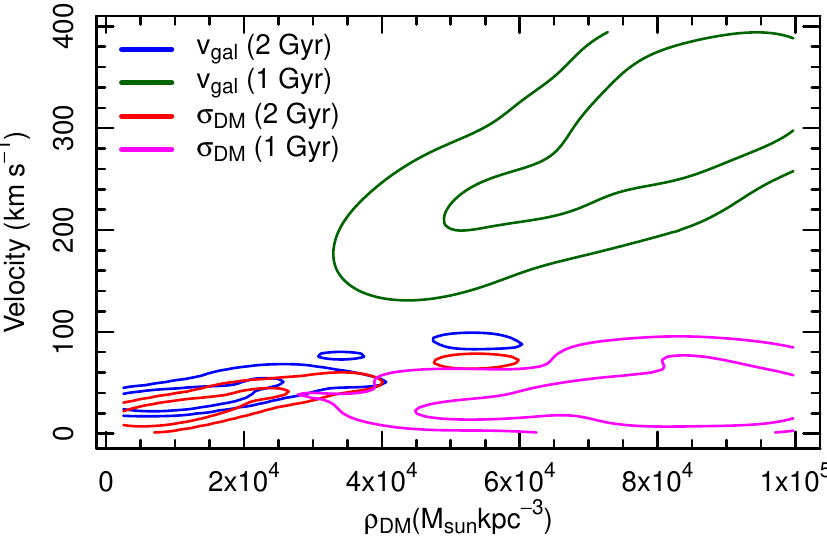}
    \caption{The environmental properties of IC 2487: dark matter density and its velocity dispersion with the amplitude of the movement of IC 2487. All of the calculations are made twice: one with orbital calculations of 1 Gyr, and one with 2 Gyrs. Substantial differences suggest that longer times require less perturbations to build up lopsidedness similar to observations. 
    }
    \label{fig:ic2487_gal_environment_properites}
\end{figure}

The velocity vector of the galaxy characterizes the relative motion between the IC 2487 galaxy and the passing particles responsible for DF and tidal effects. Passing particles might belong either to a large scale structure {(e.g. supercluster)} or to the halo of a sparse group, in the case the galaxy is a member of this group. In fact, both DM populations may contribute to generating the DM density wake responsible for DF. On the one hand, the galaxy might transfer part of its orbital energy to the DM particles of the group's halo. On the other hand, as the group moves with respect to the large scale structure, particles belonging to the latter can also induce DF. {These two populations of DM particles might have different structural properties (i.e. different DM density and velocity dispersion). Nonetheless, in our modelling the galaxy moves through a background of DM particles characterized by a unique DM density and velocity dispersion.} The absolute value of the galaxy's velocity with respect to the background of DM particles was found to be $35\pm7~{\rm km\,s^{-1}}$ and $300\pm70~{\rm km\,s^{-1}}$ when integrating orbits for $2~$Gyr and $1$~Gyr, respectively. The longer the galaxy is being perturbed, the smaller influences can be determined. Figure~\ref{fig:ic2487_gal_environment_properites} shows the two-dimensional marginalized posterior distribution for the galaxy's velocity as a function of the density of DM particles. 
Such a large variation in galaxy's speed results from the difference in the perturbation times (i.e. 1 and 2 Gyr). In principle, the time of the perturbation could be deduced by measuring peculiar line-of-sight velocities from e.g. CosmicFlows project \citep{Tully_2016}. 

\begin{figure}
    \centering
    \includegraphics{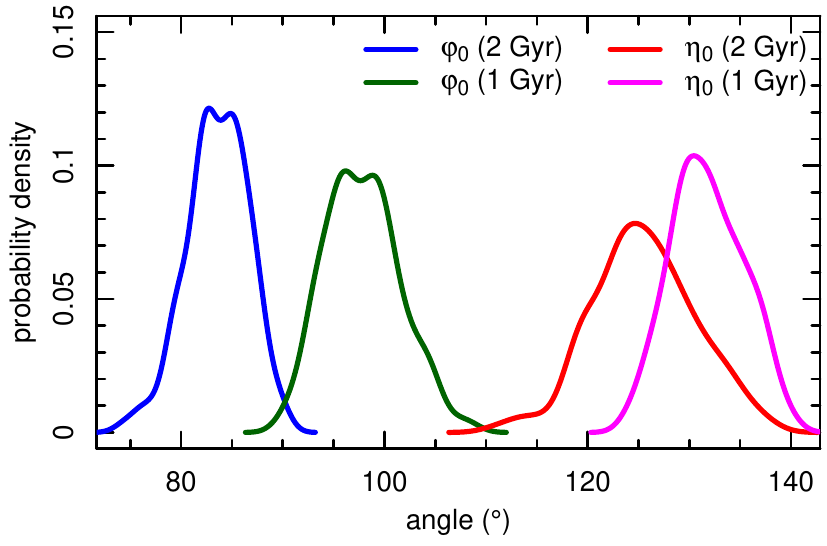}
    \caption{The direction of the galaxy described with two angles: the direction of the plane of the galaxy ($\eta_0$) and from rotation axis ($\phi_0$).     }
    \label{fig:gal_angles}
\end{figure}

The direction of the galaxy's velocity is well determined since it is more sensitive to the warp and lopsidedness directions than its absolute value. {Independently on the time of the perturbation,} the angle between the galaxy movement and the rotation axis is $123-134^\circ$, and the angle between the plane of the sky and galaxy's movement is $82-99^\circ$ if measured in the plane of the galaxy (see Fig.~\ref{fig:gal_angles}). It is to be noticed that this is not the same as the angle toward the warp in the plane as there is a phase shift between the acceleration and the warp responding to it. 

In principle, the galaxy's velocity with respect to the background of DM particles, which has been derived above, can also be found when comparing the true distance of the galaxy with its redshift (since redshift includes also peculiar velocity component) \citep{Tully_2016}. The values of the line-of-sight peculiar velocity (assuming $H_0 = 70\,{\rm km\,s^{-1}\,Mpc^{-1}}$) for the IC 2487 galaxy found in the literature range from $-347$ to $640$ ${\rm km~s^{-1}}$ \citep{Karachentsev_2006, Theureau_2007, FL:2013}. The projected velocity of IC 2487 found from our modelling is $|v_{\rm los}|\lesssim 150\,{\rm km\,s^{-1}}$. In general, these results are consistent, but unfortunately this is not very constraining to check the modelling accuracy. Although not applicable in the present case, in principle, the method provides an opportunity to mitigate the degeneracy between cosmological and Doppler redshifts. 


A more intriguing result of the present paper is the possibility to constrain the properties of the DM medium in the cosmic web and group environments. The range of the DM density is derived to lie in the range $(0.6-2.4)\times10^4~{\rm M_\odot~kpc^{-3}}$ for 2 Gyr orbit integration (this and subsequent ranges correspond to $65\%$ confidence interval in case the likelihood calculation is stable; see Sect.~\ref{sec:LLstability}), and $(5.5-9.5)\times10^4~{\rm M_\odot~kpc^{-3}}$ for $1$ Gyr orbit integration. For a comparison, the average density of the Universe is $140~{\rm M_\odot kpc^{-3}}$. The derived DM velocity dispersion lies in the ranges $20-60~{\rm km~s^{-1}}$ and $15-40~{\rm km~s^{-1}}$ for 2 Gyr and 1 Gyr, respectively. Figure  \ref{fig:ic2487_gal_environment_properites} shows the degeneracy between them and the galaxy's velocity. {The problem of not knowing the orbital integration time and the degeneracy between velocity dispersion and density can be eased by acquiring additional information of a system. For example, if a lopsided galaxy is in a cluster, its environmental density $\rho_{\rm DM}$ can be determined from external source such as weak lensing. In addition, being in a group/cluster allows to determine line of sight velocity with respect to the cluster which reduces the degeneracy from $v_{\rm gal}/\sigma_{\rm DM}$. }

\section{Discussion}\label{sec:discussion}
With this paper we tested how galactic discs would respond to the tidal forces generated by the wake also responsible for DF. Although the overall estimation of the amplitude of the effect gives reasonable results, we would like to discuss some caveats and further possibilities. 

Although we present the results for $1$ and $2$ Gyr orbit integration times, we tend to favour the latter solution. The main reason is that the solution of quicker lopsided build-up (i.e. 1 Gyr) requires that the ratio between the galaxy's velocity and DM velocity dispersion to be unreasonable in case the IC 2487 galaxy resides in a bound group. In case there is an infall to a cluster, then the galaxy velocity differs strongly from the DM particle ones and large $v_{\rm gal}/\sigma_{\rm DM}$ is expected. In that case the $1$ Gyr orbit integration solution might be viable.

\subsection{Caveats}\label{sec:caveats}
{The original cause of DF is the gravitational potential of a galaxy which is moving through a background medium of DM particles.} Our results show how this potential and {its corresponding} acceleration field are distorted. However, our modelling does not include these distortions retrospectively, i.e. {the DM wake is not affected by distortions on the gravitational potential that has initially generated it. Therefore in the regime where $a_\star$ (given by Eq.~\ref{eq:extra_acc}) is much larger than the intrinsic acceleration of the galaxy, the tidal effects might not be correctly estimated.}

The input requires gravitational potential. Initial modelling for potential includes also particles that are passing through galaxy (unbound particles) that are introduced in further modelling. Hence, some particles are counted twice or not counted at all in $a_\star$ determination.  

In the application on IC 2487, when calculating the response of the galaxy as warp and lopsidedness, the vertical acceleration component of the ${\bf a_\star}$ is adequately determined. 
The vertical restoring force for warp is evaluated from acceleration of Miyamoto-Nagai potential modelling. As there is degeneracy between inclination and thicknesses (and therefore vertical forces), the response of the disc can be sensitive to the photometric modelling, which cannot be included due to the fore-mentioned degeneracy. 

In the present paper we have assumed that the velocity distribution of DM particles is described by a normal distribution. Whether this is a good assumption for collisionless systems, in cases of some non-stationarity even more complicated distributions are possible. It was demonstrated by \citet{Leung:2020} that, at least in the case of globular clusters orbiting in a DM halo, different forms of velocity distribution of DM particles affect DF effects surprisingly significantly. Thus the assumption of the normal distribution should be handled as an approximation only.

\subsection{Outer parts of galaxies}
In order to seek the side-effects of DF, we must know where to look for. The gravitational force of the galaxy decreases with distance from the centre. Therefore, outer parts of galaxies are less strongly bound to a galaxy, and external perturbations, even with the same strength, have higher influence on the outer parts of the galaxy. This suggests that the outer parts of galaxies are more prone to be perturbed. 

We calculated that the tidal forces from the wake can take many forms, depending on the orientation between the disc's plane and the movement of the galaxy which is measured by the angle $\eta$. In Sect.~\ref{sec:magnitude_of_effect} we mentioned the possibility to produce \textit{U} and \textit{S} shaped warps as a function of $\eta$. The \textit{L} shaped warps would be produced as intermediate angles. That is, the inclination of the movement provides the diversity of the warps. {The expected shape of the warp distribution would be diverse, both in morphology and asymmetry, as shown in an example Fig.~\ref{fig:ic2487_family}. } 

Lopsided galaxies show lopsidedness mostly directed  toward one direction, without significant change of phase with radius. From the perspective of lopsidedness from DF, the same applies. The wake is located opposite to the direction of galaxy movement. A galaxy moves in a cluster or group in an orbit with a much larger period than  the galaxy's rotation period. This stability suggests that the wake is in one direction for an extended period of time, and the expected shift of the lopsidedness is minimal. This corresponds to the observations.

In the inner parts of a galaxy, tidal forces from the wake of the DF can be observed as the misalignment of the kinematic and photometric position angles, as shown in the present study (see Fig.~\ref{fig:orbit_observables}). This misalignment does not always correlate with interaction marks and satellites -- the cause of tidal forces \citep{Pilyugin_2020}. The closeness of gas and stellar kinematic alignment suggests that gravitational potential non-axisymmetry is preferred over gas accretion. The same applies to non-correlation with other properties of galaxy. Further HI surveys, such as WALLABY \citep{koribalski2020wallaby}, will shed more light on the mechanisms of the asymmetry production \citep{reynolds2020hi}.

\subsection{DF as probe for the nature of dark matter}
The observation of DF's effects will provide an additional evidence for the existence of DM in the form of particles. Furthermore, it might provide a way to prove the nature of DM. Besides gravitational interactions, additional sources of pressure and/or forces need to be taken into account in the formation of the DM density wake depending on the properties of the DM particle (e.g. self-interactions, wave-like behaviour) (see e.g. \citet{1999ApJ...513..252O, 2012JCAP...02..011L}). On top of this, the effective acceleration field acting on a galaxy depends on DM densities and velocity distributions along the galaxy orbit, which indeed differ for different DM scenarios. Therefore, the magnitude of the tidal effects correlates with the nature of the DM particle.

Although at present it is difficult to derive e.g. DM velocity distribution properties directly from observations, the modelling of tidal effects may help to constrain DM distribution properties.
However, in this case we need to assume that all or at least dominating part of tidal effects of modelled galaxies are caused by DF. This may not be the case of tight galaxy groups with frequent merger events. Most valuable for us in this sense are isolated galaxies or galaxies in loose groups where one may expect that recent mergers are insignificant and lopsidedness of galaxies is caused mainly by the DM wake responsible for DF. Constraining DM distribution properties in large scale environments of isolated galaxies and of loose groups is especially valuable as little is known about DM properties there.

In this work, we have assumed the cold dark matter (CDM) paradigm, which has been remarkably successful in explaining the large scale structure of the Universe. We leave to future work the study of the effect of DF in the density and kinematics of galaxies under the assumption of different DM scenarios.

\section{Summary}\label{sec:summary}
In the present paper we derived necessary equations and thereafter, estimated the tidal forces due to the density wake produced by dynamical friction between a galaxy and a surrounding field of dark matter particles. We showed that these tidal forces are able to produce lopsidedness and warps in galactic discs and photometric-kinematic position angle misalignments. The magnitude of these effects depend on the spatial velocity of the galaxy, its orientation with respect to the velocity direction and the distribution of dark matter particles. 

As a case study, we selected the isolated galaxy IC 2487 to estimate the properties of the background dark matter distribution in order to be able to reproduce the observed lopsidedness, warp and other HI distribution features. We estimated that the necessary local density of the surrounding dark matter field is $10^4 - 10^5\,{\rm M_\odot\,kpc^{-3}}$ and the necessary velocity dispersion of dark matter particles should be $\lesssim80\,{\rm km\,s^{-1}}$. According to large scale structure simulations the density parameter is realistic for dark matter distribution in loose groups or in overall dark matter field. As a side product, we were able to estimate the 3D velocity of the galaxy with respect to dark matter. An attempt to check the consistency of the line-of-sight velocity component with the peculiar motion from the redshift-distance differences was not successful, due to large uncertainties of distance measurements. 

Thus, at least in principle, it is possible that dynamical friction due to dark matter particles is responsible for the observed lopsidedness and warps in isolated or nearly isolated galaxies especially where other mechanisms (e.g. mergers) do not work well. 

\section{Acknowledgement}
{We are grateful to C.~Jog for refereeing the paper and providing very helpful feedback and suggestions. }
We thank Urmas Haud for useful insights on observations in the radio astronomy and Gert H\"utsi for general discussion. 
This work was supported by institutional research funding  \mbox{IUT40-2}, \mbox{PUTJD907} and \mbox{PRG803} of the Estonian Ministry of Education and Research. We acknowledge the support by the Centre of Excellence "The Dark Side of the Universe" (TK133) and MOBTP86 financed by the European Union through the European Regional Development Fund. 
Funding for the Sloan Digital Sky Survey IV has been provided by the Alfred P. Sloan Foundation, the U.S. Department of Energy Office of Science, and the Participating Institutions. SDSS-IV acknowledges support and resources from the Center for High-Performance Computing at the University of Utah. The SDSS web site is www.sdss.org.

\bibliographystyle{mnras}

\section*{Data availability}
The data underlying this article are available in the article. The sources of the code used in this paper will be shared on reasonable request to the corresponding author.
\bsp
\label{lastpage}
\end{document}